\documentstyle[12pt,epsf]{article}
\newcommand{\be}{\begin{equation}}
\newcommand{\ee}{\end{equation}}
\newcommand{\ba}{\begin{eqnarray}}
\newcommand{\ea}{\end{eqnarray}}
\newcommand{\bb}{\begin{thebibliography}}
\newcommand{\eb}{\end{thebibliography}}
\newcommand{\ci}[1]{\cite{#1}}
\newcommand{\bi}[1]{\bibitem{#1}}
\newcommand{\lab}[1]{\label{#1}}

\begin{document}
%\phantom{.}
%\vspace{-2cm}
\begin{center}
{\large{\bf Spin asymmetries in diffractive high-energy
reactions.\footnote{Talk presented at the Adriatico Research Conference
"Trends in Collider Spin Physics" (5 - 8 December 1995) ICTP, Miramare
Trieste, Italy} }} \\ S.V.Goloskokov \footnote{Email:
goloskkv@thsun1.jinr.dubna.su}\\ Bogoliubov Laboratory of Theoretical
  Physics, Joint Institute for\\  Nuclear Research, Dubna 141980, Moscow
  region, Russia.

\end{center}

\vspace{.2cm}
\begin{abstract}
The analysis of some effects caused by the  spin--dependent pomeron
couplings is presented. It is shown that the structure of pomeron--proton
and quark--pomeron couplings can be tested in future polarized
experiments on elastic $pp$ reactions and diffractive $Q \bar Q$
production.
\end{abstract}

 Extensive polarized programs are proposed at HERA, RHIC and LHC
accelerators (see e.g.\ci{now,bunce,gur}).  Among different processes which
can be studied there are the diffractive and elastic high energy reactions
predominated by the pomeron exchange.  The study of the pomeron properties
is a very popular problem now because of the observation of events with a
large rapidity gap at CERN \ci{ua8} and DESY \ci{h1}. These events may be
caused by the diffractive reactions that can be investigated by using the
QCD--models for the pomeron.

For the diffractive scattering of polarized particles, the question
about the spin structure of the pomeron appears. This problem is very
important for the following reasons:
\begin{itemize}
\item{There are many observations of spin effects at high energies and
fixed momenta transfer \ci{nur}.}
\item{Some model approaches predict nonzero spin effects in the $s \to
\infty, |t|/s \to 0 $ limit (see \ci{gol,soff} e.g.).}
\item{Attempts to extract the spin-flip amplitude from the experimental
data \ci{akch} show that the ratio of spin-flip and spin-non-flip
amplitudes can be not small and independent of energy.}
  \end{itemize}
Just in all of these cases the pomeron exchange should contribute.
So, there is a possibility that the pomeron has a complicated spin structure.

The high-energy two-particle amplitude determined by the pomeron
exchange can be written in the form
\be
T(s,t)=i I\hspace{-1.6mm}P(s,t) V_{h_1h_1I\hspace{-1.1mm}P}^{\mu} \otimes
V^{h_2h_2 I\hspace{-1.1mm}P}_{\mu}.    \lab{tpom}
\ee
Here  $I\hspace{-1.6mm}P$ is a function caused by the pomeron,
$V_{\mu}^{hhI\hspace{-1.1mm}P}$ are the pomeron-hadron vertices.
The calculation of this amplitude in the nonperturbative two-gluon exchange
model  \ci{la-na} and in the
BFKL model \ci{bfkl} shows that the pomeron couplings are simple in form
(the standard coupling in what follows):
\be
V^{\mu}_{hh I\hspace{-1.1mm}P} =\beta_{hh I\hspace{-1.1mm}P}\; \gamma^{\mu},
\lab{pmu}
\ee
In this case the spin-flip effects are suppressed as a power of $s$.

The situation does change drastically when the large-distance loop
contributions are considered that complicate the spin structure
of the pomeron coupling. These effects can be determined by the hadron wave
function for the pomeron-hadron couplings or by the gluon-loop $\alpha_s$
corrections for the quark-pomeron coupling.  As a result, the spin
asymmetries appear that have weak energy dependences as $s \to \infty$.

The main purpose of this report is to study the single transverse spin
asymmetry in diffractive reactions.
Note that the common perturbative QCD approach to the single--spin
asymmetry was proposed in \ci{ter}.

The single spin asymmetry depends strongly on
the hadron properties. It is determined by the relation
\be
A_{\perp}=\frac{\sigma(^{\uparrow})-\sigma(^{\downarrow})}
{\sigma(^{\uparrow})+\sigma(^{\downarrow})}=
\frac{\Delta \sigma}{\sigma} \propto
 \frac{\Im (f_{+}^{*} f_{-})}{|f_{+}|^2 +|f_{-}|^2},  \lab{astr}
\ee
where $f_{+}$ and $f_{-}$ are spin-non-flip and spin-flip amplitudes,
respectively. So, single spin asymmetry appears if both $f_{+}$ and $f_{-}$
are nonzero and there is a phase shift between these amplitudes.
We shall discuss some consequences of the new spin-dependent form
of the pomeron vertices in elastic scattering and diffractive $Q \bar Q$
production that can be studied in future polarized ezperiments at
HERA, RHIC and LHC.\\[0.2cm]

{\large{\bf Pomeron-proton vertex effects }}\\

Pomeron-proton coupling is connected mainly with the proton structure at
large distances.  This coupling determines the  single ($A_{\perp}$) and
double ($A_{nn}$) transverse spin asymmetries at high energies and fixed
momenta transfer.

The perturbative calculation of this coupling is rather difficult.
Moreover, for a momentum transfer about few $GeV^2$ the nonperturbative
contributions should be important. One of the models that takes into
account these effects is the diquark model \ci{kroll}. This model
can be used to study the spin structure of the pomeron-proton
coupling.

In this part of the report we shall discuss  the predictions of the
meson-cloud model \ci{gol} obtained in collaboration with O.Selyugin.
This model effectively considers the proton structure at large distances.
It leads to the following form of the pomeron-proton coupling
\be
V_{ppI\hspace{-1.1mm}P}^{\mu}(p,r)=m p_{\mu} A(r)+ \gamma_{\mu} B(r),
\lab{prver}
\ee
where $m$ is the proton mass and $r$ is the momentum transfer ($t=r^2$).
Here $ \gamma_{\mu} B(r)$ is a standard pomeron coupling like (\ref{pmu})
that determines the spin-non-flip amplitude. The term $m p_{\mu} A(r)$ is
caused by the meson-cloud effects. The coupling (\ref{prver}) leads to the
spin-flip in the pomeron vertex that does not vanish in the $s \to \infty$
limit. Really, using the vertex (\ref{prver}) we can estimate the
spin-non-flip and spin-flip effects from the pomeron-proton vertex \ba
|f_{+}(s,t)| \propto s \;|B(r)|; \nonumber\\
|f_{-}(s,t)| \propto m \; \sqrt{|t|}\; s\; |A(r)|. \lab{fpm}
\ea
So, both the amplitudes have the same energy dependence.
The model predicts the following ratio for spin-flip
and non-flip amplitudes
\be
\frac{\;\;\;m\;|f_{-}(s,t)|}{\sqrt{|t|}\;|f_{+}(s,t)|} \simeq
\frac{\;\;\;m^2\;|A(r)|}{|B(r)|} \simeq
  0.05 \div
0.07 \;\;\; {\rm for} \;\;\;  |t| \sim 0.5GeV^2  \lab{fr}
  \ee
that is consistent with estimations of ref. \ci{akch}.

In the model \ci{gol} the amplitudes $A$ and $B$ have a phase shift caused
by the soft pomeron rescattering effect. As a result, the single-spin
asymmetry determined by the pomeron exchange
\be
A_{\perp} \simeq \frac{2m \sqrt{|t|} \Im (AB^{*})}{|B|^2} \lab{epol}
\ee
appears. We omit here the $|A|^2$ term in the denominator because its
contribution is rather small (\ref{fr}). The asymmetry (\ref{epol}) has a
weak energy dependence.

The meson-cloud model \ci{gol} and the model of a rotating matter current
 \ci{soff} describe all known experimental data on elastic $pp$ scattering
 at fixed momenta transfer quantitatively and can predict the physical
 observables (cross-sections, asymmetries) at higher energies. The
 predictions of the model \ci{gol} for the differential cross section,
 transverse single spin and double spin asymmetries for the RHIC energy
(LHS fixed target experiment energy) $\sqrt{s}=120GeV$ are shown in Figs
1-3.  The model \ci{soff} predicts a similar absolute value for
$A_{\perp}$ asymmetry but it is opposite in sign near the diffraction
minimum. So, the future PP2PP experiment proposed at RHIC \ci{gur} can
give an important information about the mechanism of spin--effect
generation at the pomeron-proton vertex.

Moreover, the future PP2PP experiment at RHIC will give a possibility to
measure the spin-dependent cross-section with parallel $\sigma(\uparrow
\uparrow)$ and antiparallel $\sigma(\uparrow \downarrow)$ polarization in
proton--proton scattering.  It can be shown that in this case the
energy dependence of the spin--flip  and
spin--non--flip amplitude can be measured.  Really, using the standard
notation for the $pp$ helicity amplitudes, let us suppose that the
double-spin-flip amplitudes are small with respect to the spin--non--flip
one $F_2(s,t) \sim F_4(s,t) \ll F_1(s,t)$ and spin-non-flip amplitudes are
approximately equal $F_{++}(s,t)=F_1(s,t) \sim F_3(s,t)$. These usual
restrictions are true for the models \ci{gol,soff}.  In this case the
observables are determined by two  amplitudes $F_{++}(s,t)$ and
$F_{+-}(s,t)=F_5(s,t)$ and  we can find
\ba
|F_{+-}(s,t)|^2  \propto (\sigma(\uparrow \uparrow)-
\sigma(\uparrow\downarrow)) \propto A_{nn}(\sigma(\uparrow \uparrow)+
\sigma(\uparrow \downarrow)) \nonumber \\
|F_{++}(s,t)|^2 \propto \sigma(\uparrow\downarrow) \propto
(1-A_{nn})(\sigma(\uparrow \uparrow)+ \sigma(\uparrow \downarrow)) .
 \ea
The energy dependence of $F_{++}(s,t)$ and $F_{+-}(s,t)$ amplitudes
coincide with (\ref{fpm}) for the $V_{ppI\hspace{-1.1mm}P}$ vertex
(\ref{prver}).  So, we shall have the same energy dependence of
$(\sigma(\uparrow \uparrow)- \sigma(\uparrow\downarrow))$ and
$\sigma(\uparrow\downarrow)$ for the spin--dependent pomeron--proton
vertex and the ratio of these quantities will be practically energy--
independent (Fig.4).  Otherwise, we shall find the rapid decrease of the
ratio
$$(\sigma(\uparrow \uparrow)-\sigma(\uparrow\downarrow))/
\sigma(\uparrow\downarrow) \propto 1/s$$
 with growing $s$ (see Fig.4) caused by the
standard energy dependence of the spin-flip amplitude.
The low-energy experimental point for the ratio
$(\sigma(\uparrow \uparrow)-\sigma(\uparrow\downarrow))/
\sigma(\uparrow\downarrow)$ from \ci{kr} is shown in Fig.4 too. This point
does not contradict the weak energy dependence of this ratio predicted
by the model \ci{gol}. The energy dependence of the cross sections
$\sigma(\uparrow \uparrow)$ and $\sigma(\uparrow \downarrow)$ can be
studied experimentally at RHIC.  As a result, the direct information about
 the spin--flip effects in the pomeron--proton coupling
 should be obtained.  \\ [0.2cm]

{\large{\bf Quark-pomeron vertex effects}} \\

The standard quark--pomeron coupling (\ref{pmu}) is
determined by the contributions where two gluons from the pomeron
interact with a single quark in the hadron. The nonplanar
graphs in which gluons from the pomeron interact with different quarks
in the loop, as a rule, do not exceed $10$ percent as compared
to the planar--diagram contributions at fixed momenta  transfer (see e.g.
\ci{goljp}). The large--distance gluon--loop corrections to the
quark--pomeron coupling have been studied in \ci{gol-pl}. It is shown that
they lead to new spin structures of the quark--pomeron vertex.  The
perturbative calculations \ci{gol-pl} give the following
form for this vertex:
\be
V_{qqI\hspace{-1.1mm}P}^{\mu}(k,r)=\gamma^{\mu} u_0+2 M_Q k^{\mu} u_1+
2 k^{\mu}
/ \hspace{-2.3mm} k u_2 + i u_3 \epsilon^{\mu\alpha\beta\rho}
k_\alpha r_\beta \gamma_\rho \gamma_5+i M_Q u_4
\sigma^{\mu\alpha} r_\alpha,    \label{ver}
\ee
where $k$ is the quark momentum, $r$ is the momentum transfer and  $M_Q$
is the quark mass.  So, in addition to the $\gamma_\mu$ term, the new
structures immediately appear from the loop diagrams.  The functions
$u_1(r) \div u_4(r)$ are proportional to $\alpha_s$. It has been shown
\cite{gol4} that these
functions can reach $30 \div 40 \%$ of the standard pomeron term
$u_0(r)$
for $|r^2| \simeq {\rm Few}~ GeV^2$. Note that
the spin structure of the quark-pomeron coupling (\ref{ver}) is
drastically different from the standard one (\ref{pmu}).
Really, the terms
$u_1(r)-u_4(r)$ lead to the spin-flip at the quark-pomeron vertex in
contrast
with the term $u_0(r) \gamma_\mu$.

This new complicated form of the pomeron--quark coupling (\ref{ver})
should modify various  spin asymmetries in high--energy diffractive
reactions \cite{klen,golasy}.
It has been shown that a simplest way to test the quark--pomeron
coupling is to study the single pomeron two--jet production in
lepton--proton and proton--proton reactions.

The double spin longitudinal asymmetries has been investigated in
\cite{golasy}.  It has been shown that these asymmetries in hadron--hadron
and lepton--hadron reactions do not depend practically on the
pomeron--proton coupling structure. They are sensitive to the form of
quark--pomeron coupling, especially to the terms $u_0(r)$ and $u_3(r)$ in
(\ref{ver}).

Here we shall analyse the single--spin transverse asymmetry in diffractive
$Q \bar Q$ production.  In the  diffractive jet production at small
$x_p$ (fraction of the initial proton momentum carried off by the
pomeron) the main contribution is determined by the region where the
quarks in the loop are not far of the mass shell. In this case we can use
here the same  "soft pomeron" \cite{softp} as for the elastic
 reactions.  As a result, in the asymmetry of diffractive $Q \bar Q$
production the same single-spin hadron asymmetry (\ref{epol}) determined
by the  "soft pomeron" as in the case of elastic scattering should appear.
In our further estimations we shall use the magnitude $A^h_{\perp}=0.1$
that is consistent with the experiment and the model \ci{gol} results.

The cross sections $\sigma$ and $\Delta \sigma $ determined
in (\ref{astr})
can be written in the form
\begin{equation}
\frac{d \sigma(\Delta \sigma)}{dx_p dt dk_{\perp}^2}=\{1,A^h_{\perp}\}
\frac{\beta^4 |F_p(t)|^2 \alpha_s}{128 \pi s x_p^2}
\int_{4k_{\perp}^2/sx_p}^{1}
\frac{dy g(y)}{\sqrt{1-4k_{\perp}^2/syx_p}}
\frac{ N^{\sigma(\Delta \sigma)}
(x_p,k_{\perp}^2,u_i,|t|)}{(k_{\perp}^2+M_Q^2)^2}. \label{si}
\end{equation}
Here $g$ is the gluon structure function
of the proton, $k_{\perp}$ is the transverse momentum of jets, $M_Q$
is the quark mass, $N^{\sigma(\Delta \sigma)}$ is the
trace over the quark loop,
$\beta$ is the pomeron coupling constant, $F_p$
is the pomeron-proton form factor. All the information on the quark-pomeron
vertex structure is concentrated in $N^{\sigma}$ and
$N^{\Delta \sigma}$ functions.

It has been found that the main contributions to $N^\sigma(N^{\Delta
\sigma})$ in the discussed region come mainly from $u_0$ and $u_3$
structures in (\ref{ver}). Moreover, the ratio $N^{\Delta \sigma}/N^\sigma
\simeq 0.5$ and is independent of $k_{\perp}^2$ for the standard
quark--pomeron vertex but the same ratio is the $k_{\perp}^2$ dependent
for the pomeron coupling (\ref{ver}).  Our predictions for single spin
asymmetry for the  proposed HERA-N experiment \ci{now} can be found in
\ci{golhera,goltr}.

Here we would like to show our results for $\sigma$ and single spin
asymmetry (Figs. 5, 6) for the light--quark jet production
at the RHIC energy (LHS fixed target experiment energy)
$\sqrt{s}=120GeV$, $x_p=0.05$ and $|t|=1GeV^2$. It is easy to see
that the shape of asymmetry is different for the standard (\ref{pmu}) and
spin-dependent pomeron vertex (\ref{ver}). In the first case it is
approximately constant, in the second it increases with
$k_{\perp}^2$.  So, this asymmetry can be used for the study of the
quark--pomeron vertex structure.

The  cross sections $\sigma$ and
$\Delta \sigma$ integrated over $k^2_{\perp}$ of jets have been calculated
too. It was found that the asymmetry determined from the integrated
cross sections does not practically depend on the quark--pomeron vertex
structure. It can be written in any case  in the form
\begin{equation}
A1=\frac{\int dk^2_{\perp} \Delta  \sigma}{\int dk^2_{ \perp}\sigma}
\simeq 0.5 A^h_{\perp}  \label{a1}
\end{equation}
Thus, we can conclude that the integrated asymmetry (\ref{a1}) can be used
for studying
the hadron asymmetry $A^h_{\perp}$ caused by the pomeron.

We have presented here the analysis of some effects of the
spin--dependent pomeron couplings. It is found that the structure of these
couplings can be tested in elastic $pp$ reactions and diffractive
$Q \bar Q$  production.

{\bf The information about the pomeron--proton vertex structure can be
obtained from:}
\begin{itemize} \item{Single transverse spin asymmetry in
elastic reactions and diffractive $Q \bar Q$  production.}
\item{Double--spin $A_{nn}$ asymmetry in elastic reactions.}
\item{Energy dependence of the ratio  $(\sigma(\uparrow
\uparrow)-\sigma(\uparrow\downarrow))/ \sigma(\uparrow\downarrow)$  in
elastic reactions.}
\end{itemize}

{\bf The properties of quark--pomeron vertex can be studied from:}
\begin{itemize}
\item{$k_{\perp}^2$--dependence of the single transverse spin asymmetry in
two--jet production determined by the single--pomeron exchange in $pp$
reactions.}
\item{$A_{ll}$ longitudinal spin asymmetry in  diffractive $Q \bar Q$
production in lepton--proton and proton--proton reactions.}
\end{itemize}

Note that the spin--structure of the pomeron couplings are determined by
the large--distance gluon-loop correction or by the effects of the hadron
wave function. So, the important test of the spin structure of QCD at
large distances can be carried out by the study of diffractive
reactions in future polarized experiments at HERA, RHIC and LHC
accelerators.

 This work was supported in part by the Russian Foundation for
Fundamental Research, Grant 94-02-04616.
%\newpage

\newpage
  \vspace*{-1.5cm}
\epsfxsize=14cm
\centerline{\epsfbox{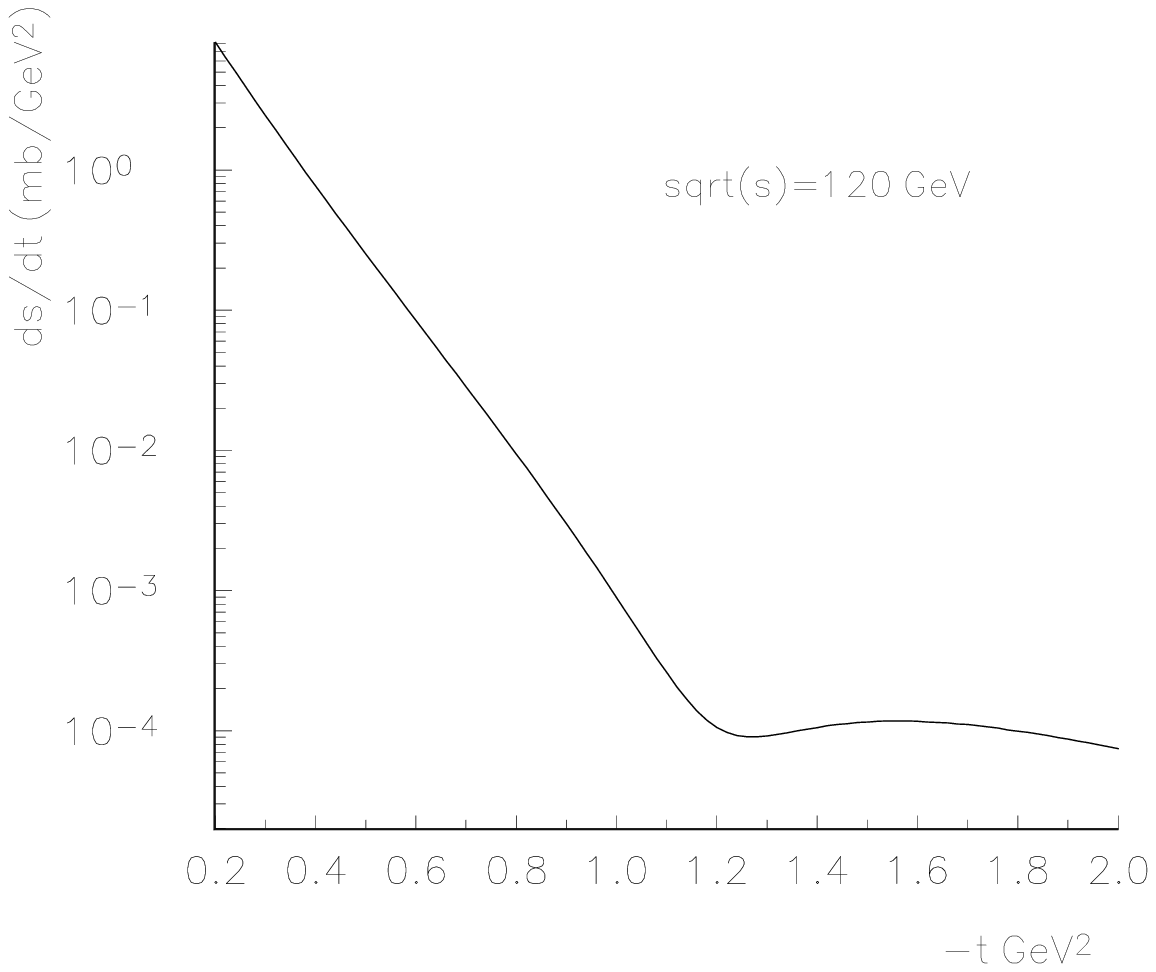}}
  \vspace*{-.2cm}
\centerline{Fig.1 ~Predictions for differential cross section
of $pp$ scattering.}

  \vspace*{-11.5cm}
 \hspace*{1.cm}
\epsfxsize=18cm
\centerline{\epsfbox{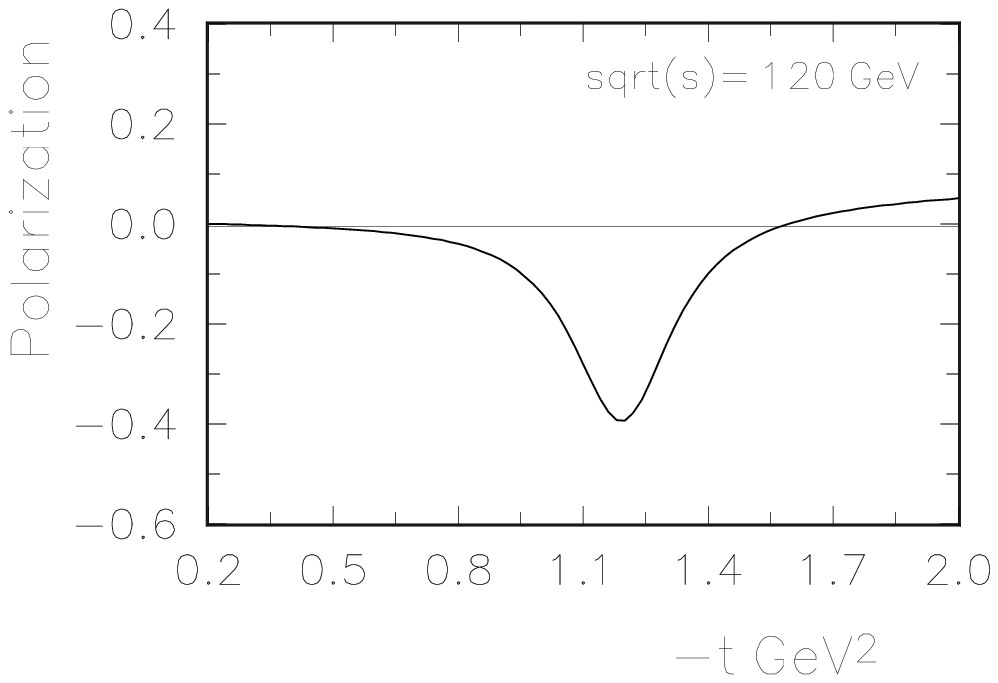}}
  \vspace*{-.2cm}

\centerline{Fig.2 ~Predictions for single-spin transverse asymmetry
of $pp$ scattering.}
\newpage
  \vspace*{-13.5cm}
 \hspace*{1.cm}
 \epsfxsize=18cm
\centerline{\epsfbox{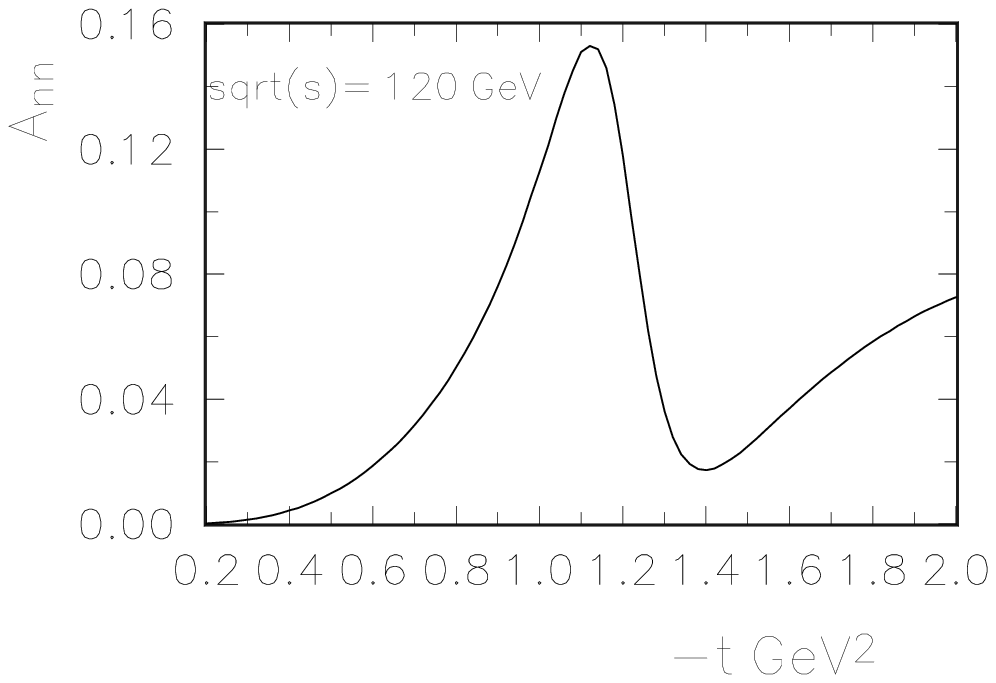}}
  \vspace*{-.2cm}
\centerline{Fig.3 ~Predictions for double-spin transverse asymmetry of
 $pp$ scattering.}

  \vspace*{2.5cm}
 \hspace*{-1.cm}
\epsfxsize=10cm
\centerline{\epsfbox{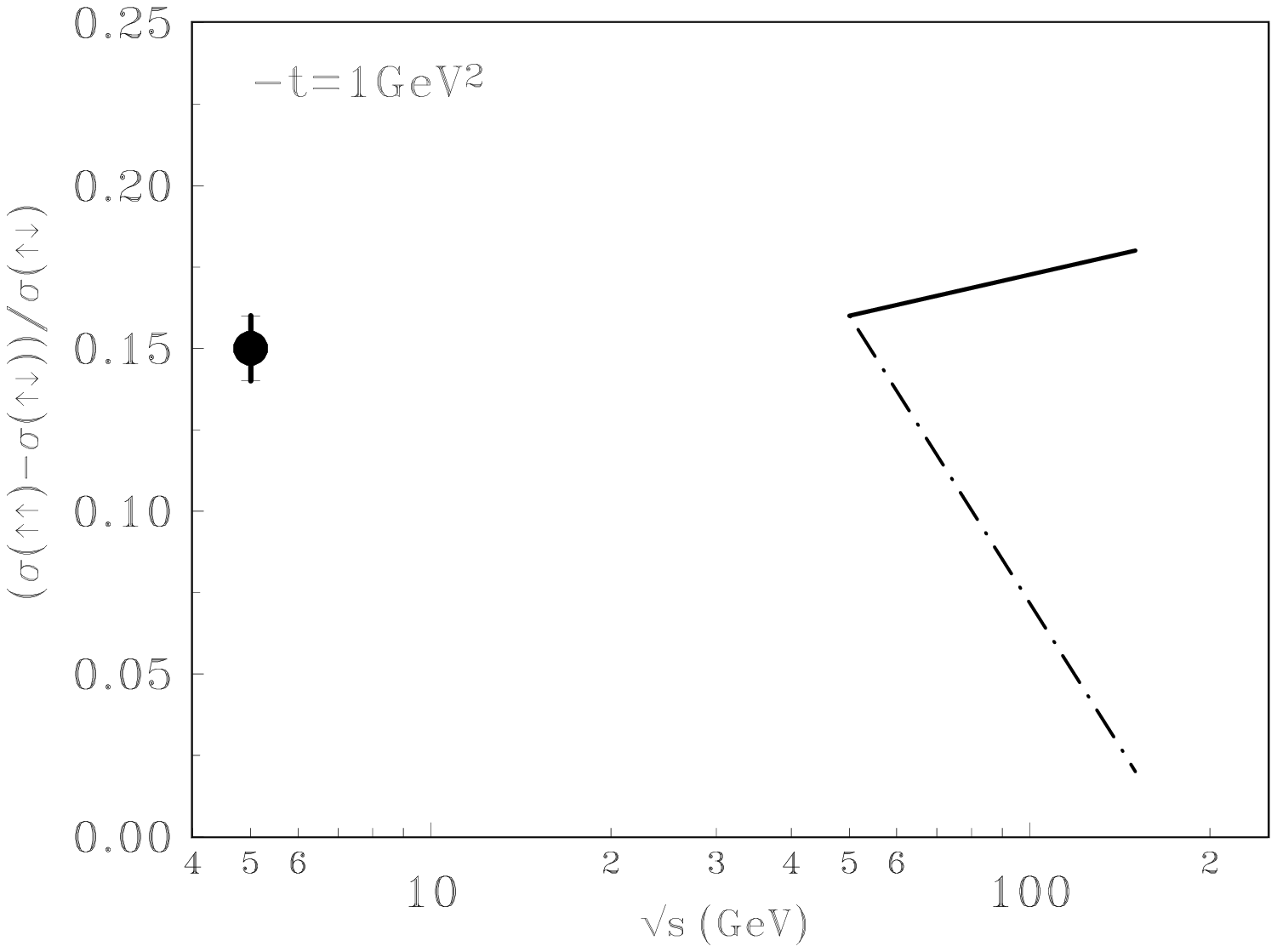}}
  \vspace*{.2cm}

Fig.4 ~Model predictions for the ratio of spin-dependent cross section for
 $pp$ scattering:
solid line -for spin-dependent pomeron vertex:
dot-dashed line -for the standard energy behaviour of the spin-flip
amplitude.
Experimental point is taken from [14].
\newpage
  \vspace*{1cm}
 \epsfxsize=10cm
\centerline{\epsfbox{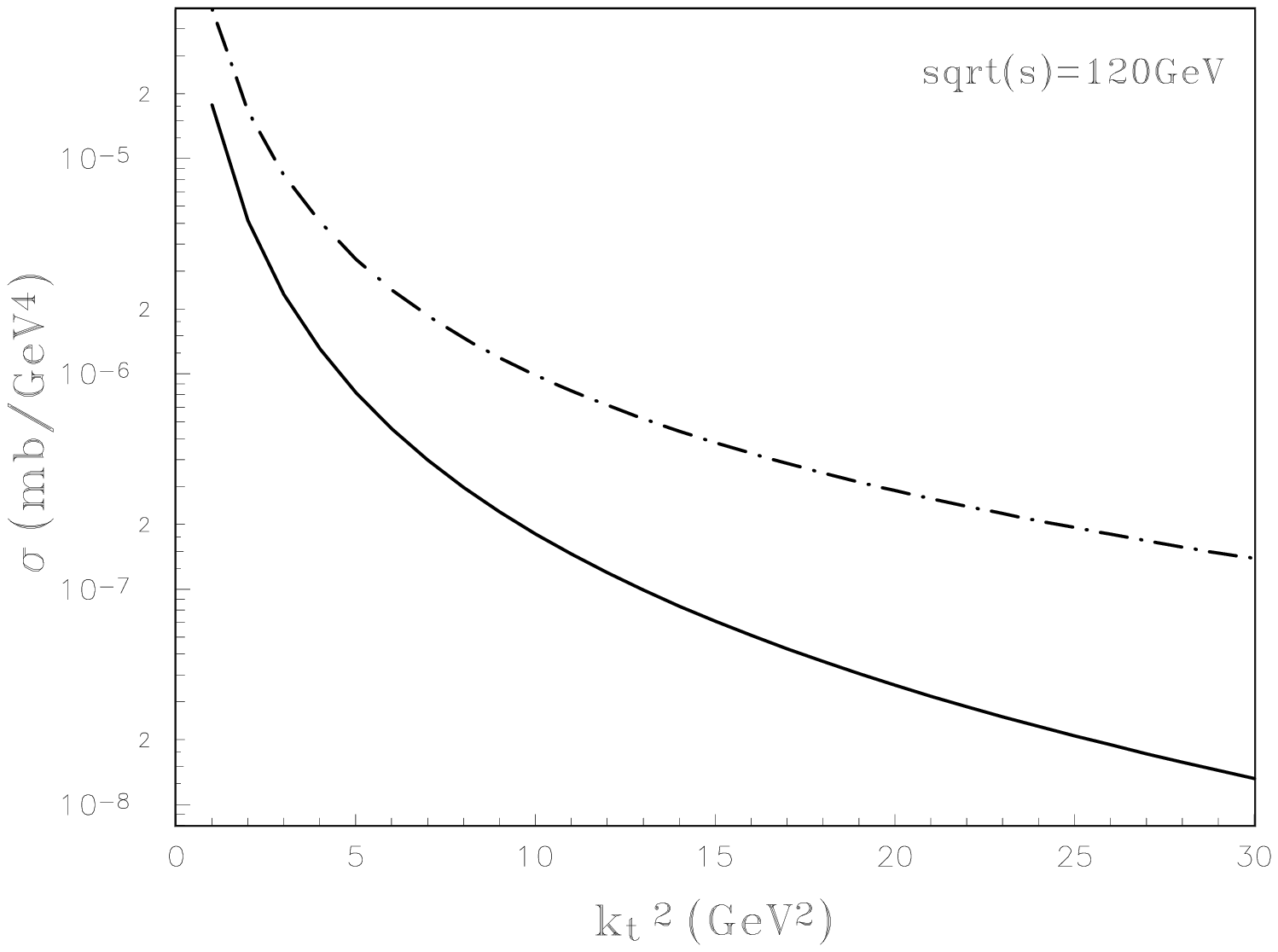}}
  \vspace*{-.2cm}
Fig.5
~Distribution of $\sigma$ over jets $k_{\perp}^2$
in diffractive light-quark production:
solid line -for standard vertex;
dot-dashed line -for spin-dependent quark-pomeron vertex.

  \vspace*{1.5cm}
% \hspace*{-1.cm}
\epsfxsize=10cm
\centerline{\epsfbox{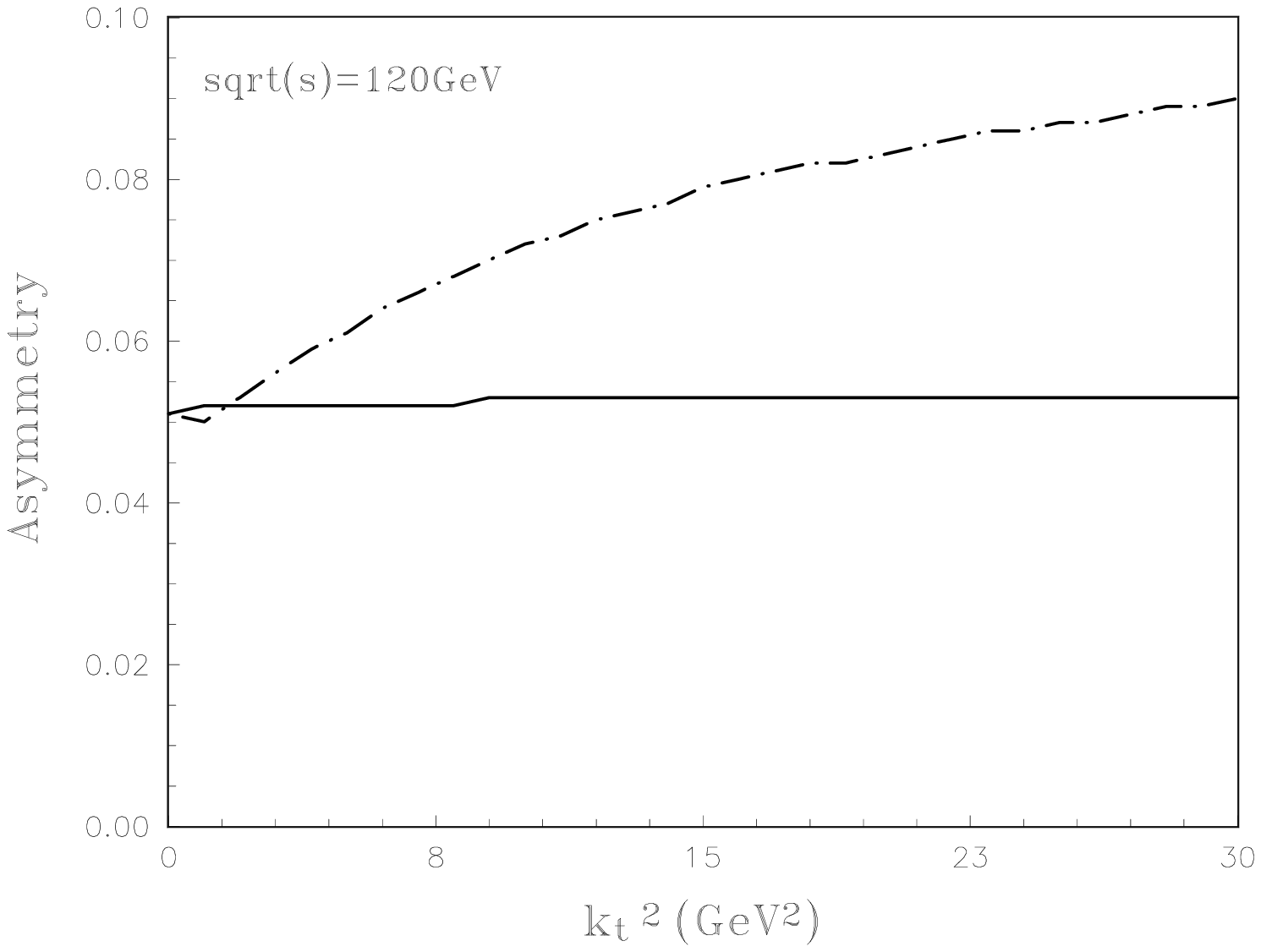}}
  \vspace*{.2cm}

Fig.6 ~ $k_{\perp}^2$ dependence of single-spin-asymmetry
in diffractive light-quark production:
solid line -for standard vertex;
dot-dashed line -for spin-dependent quark-pomeron vertex.

\end{document}